\documentclass[aps,prd,notitlepage,nofootinbib,superscriptaddress,twocolumn,floatfix]{revtex4-2}

\pdfoutput=1

\usepackage{graphicx}
\usepackage{soul}
\usepackage{subfigure}
\usepackage{latexsym}
\usepackage{amssymb}
\usepackage{amsmath}
\usepackage{color}
\usepackage{array}
\usepackage{mathrsfs}
\usepackage{xparse}
\usepackage{bbding}
\usepackage{enumitem}
\usepackage{pifont}
\usepackage{enumerate} 
\usepackage{color}
\usepackage{mathrsfs}
\usepackage{xcolor}
\usepackage{changes}
\usepackage{comment}
\usepackage[colorlinks=true,linkcolor=blue,urlcolor=blue,citecolor=blue]{hyperref}

\begin{document}

%%%%%%%%%%%%%%%%%%%%%%%%%%%%%%%%%%%%%%%%%%%%%%%%%%%%%%%%%%%%%%%
 \newcommand{\bq}{\begin{equation}}
 \newcommand{\eq}{\end{equation}}
 \newcommand{\bqn}{\begin{eqnarray}}
 \newcommand{\eqn}{\end{eqnarray}}
 \newcommand{\nb}{\nonumber}
 \newcommand{\lb}{\label}
 
\newcommand{\La}{\Lambda}
\newcommand{\va}{\scriptscriptstyle}
\newcommand{\be}{\nopagebreak[3]\begin{equation}}
\newcommand{\ee}{\end{equation}}

\newcommand{\ba}{\nopagebreak[3]\begin{eqnarray}}
\newcommand{\ea}{\end{eqnarray}}

\newcommand{\la}{\label}
\newcommand{\n}{\nonumber}
\newcommand{\su}{\mathfrak{su}}
\newcommand{\SU}{\mathrm{SU}}
\newcommand{\U}{\mathrm{U}}

\def\be{\nopagebreak[3]\begin{equation}}
\def\ee{\end{equation}}
\def\ba{\nopagebreak[3]\begin{eqnarray}}
\def\ea{\end{eqnarray}}
\newcommand{\f}{\frac}
\def\rmd{\rm d}
\def\lp{\ell_{\rm Pl}}
\def\d{{\rm d}}
\def\fe{\mathring{e}^{\,i}_a}
\def\fw{\mathring{\omega}^{\,a}_i}
\def\fq{\mathring{q}_{ab}}
\def\t{\tilde}

\newcommand{\LG}[1]{\textcolor{orange}{[{\bf LG}: #1]}}

\title { Genericness of quantum damping of cosmological shear in modified loop quantum cosmology}

\author{Wen-Cong Gan}
\email{ganwencong@jxnu.edu.cn}
\affiliation{College of Physics and Communication Electronics, Jiangxi Normal University, Nanchang, 330022, China}

\author{Leila L. Graef}
\email{leilagraef@id.uff.br} 
\affiliation{Instituto de F\'isica, Universidade Federal Fluminense, 24210-346 Niteroi, RJ, Brazil}

\author{Rudnei O. Ramos}
\email{rudnei@uerj.br}
\affiliation{Departamento de F\'isica Te\'orica, Universidade do Estado do Rio de Janeiro, 20550-013 Rio de Janeiro, RJ, Brazil}

\author{Gustavo S. Vicente}
\email{gustavo@fat.uerj.br}
\affiliation{Faculdade de Tecnologia, Universidade do Estado do Rio de Janeiro, 27537-000 Resende, RJ, Brazil}

\author{Anzhong Wang \thanks{Corresponding author}}
\email{Anzhong$\_$Wang@baylor.edu; the corresponding author}
\affiliation{GCAP-CASPER,  Department of Physics $\&$ Astronomy, Baylor University, Waco, Texas 76798-7316, USA}

%%%%%%%%%%%%%%%%%%%%%%%%%%%%%%%%%%%%%%
\begin{abstract}

In arXiv:2603.18175, the authors argue, based on numerical studies of particular cases, that the quantum damping of cosmological shear in a modified loop quantum cosmological model (mLQC-I) that was recently found in arXiv:2510.14021 is not generic and that the universe never becomes truly classical. 
In this brief Note, we revisit these claims by carefully examining the underlying assumptions and the class of initial conditions considered. We show that the examples analyzed in arXiv:2603.18175 correspond to configurations that do not represent physically admissible collapsing Bianchi I universes, as they involve mixed expanding–contracting directions and lead to effectively lower-dimensional post-bounce geometries. Restricting to physically relevant initial conditions corresponding to genuine three-dimensional contraction, we find that the quantum damping of cosmological shear is a robust dynamical feature.
 This conclusion is supported by both numerical and  perturbative analyses, which demonstrate that the post-bounce evolution admits an isotropic attractor, with anisotropies decaying exponentially and independently of the matter content, provided that the weak energy condition is satisfied. We further outline a plausible post-bounce mechanism for the onset of classicalization.

\end{abstract}
%%%%%%%%%%%%%%%%%%%%%%%%%%%%%%%%%%%%%%%%

\maketitle
 
%%%%%%%%%%%%%%%%%%%%%%%%%%%%%%%%%%%%%%%%
\section{Introduction}

Realizing a viable cosmological bounce \cite{Khoury:2001wf,Khoury:2003rt,Finelli:2001sr,Lehners:2008vx,Battefeld:2014uga,Brandenberger:2016vhg,Ijjas:2024oqn,Ashtekar:2011ni,Li:2021mop,Agullo:2023rqq,Li:2023dwy} requires overcoming a well-known difficulty: the growth of anisotropies during the contracting phase. In anisotropic spacetimes, the shear grows faster than standard matter or radiation, effectively behaving as a stiff component. Unless this growth is sufficiently suppressed, anisotropies can dominate the pre-bounce dynamics, leave large post-bounce imprints incompatible with observations, and in some cases even  preventing the bounce from taking place. Understanding how anisotropies can be controlled across the bounce therefore remains a central open problem.

In \cite{Gan:2025uvt}, we addressed this  {important issue} within a  realization of modified loop quantum cosmology (mLQC-I), in which the quantization of the Hamiltonian more closely reflects the full structure of loop quantum gravity. We showed that the shear is dynamically suppressed purely due to quantum geometric effects and decays exponentially within the quantum regime right after the bounce, independently of matter content  {satisfying the weak energy condition}. %$\omega > -1$. 
The dynamics therefore drive the universe naturally toward a homogeneous and isotropic expanding phase, without fine-tuning. This suggests that mLQC-I provides a robust mechanism for post-bounce isotropization without provoking other mechanisms, such as introducing an 
ekpyrotic phase near the bounce, during which the potential of the scalar field becomes negative, so its effective equation of  state is greater than 1, whereby the evolution of the universe is dominated by the scalar field, instead of the shear. As a result, the anisotropy problem is resolved \cite{Khoury:2001wf,Khoury:2003rt}. This mechanism works well in ekpyrotic and matter bounce models, as well as in their extended versions \cite{Lehners:2008vx,Battefeld:2014uga,Brandenberger:2016vhg,Ijjas:2024oqn}, because an inflationary phase in such models is usually absent, and  scale-invariant perturbations can be generated during the contracting phase \cite{Wands:1998yp}. 

However, in loop quantum cosmology (LQC)~\cite{Ashtekar:2011ni,Agullo:2023rqq,Li:2023dwy} and its modified versions~\cite{Li:2021mop} not only the big bang singularity is generically resolved, but also an inflationary phase in the post-bounce region generically occurs~\cite{Ashtekar:2011rm,Li:2019ipm}. Therefore, LQC and mLQCs have the advantages of both inflation (in which the problems of the Big Bang cosmology~\cite{Baumann:2009ds} are resolved, and meanwhile   scale-invariant perturbations are created) and ekpyrotic/matter bounce models (in which the Big Bang singularity is resolved). However, recently it was found that the presence of an ekpyrotic phase  can severely restrict the region of initial conditions leading to successful 
inflation \cite{Brown:2025hcb} in LQC and mLQC-I. Therefore, the mechanism found in  \cite{Gan:2025uvt} is important, as it is a built-in mechanism  
without  provoking any additional ones to suppress the shear. As a result,    
 the naturalness of post-bounce inflation in mLQC-I persists, even if large shear exists  in the contracting phase.

%%%%%%%%%%%%%%%%%%%%%%%%%%%%%%%%%%%%%%%%%%
\begin{figure}[h!]  
\subfigure[]
{\includegraphics[width=0.47\textwidth]{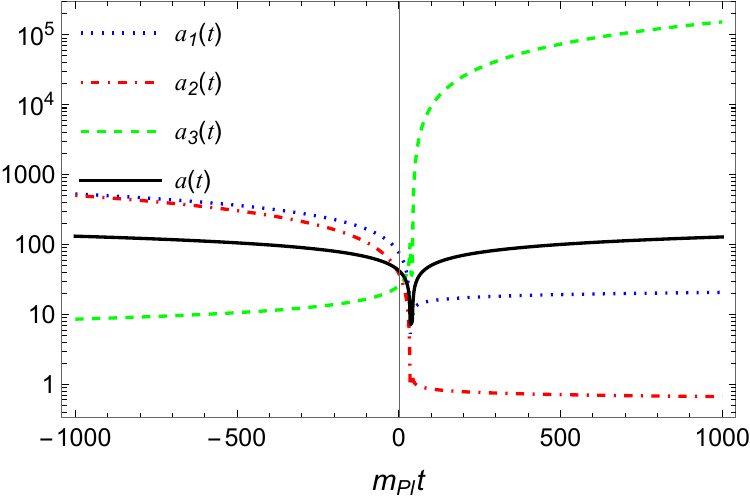}}
\subfigure[]
{\includegraphics[width=0.47\textwidth]{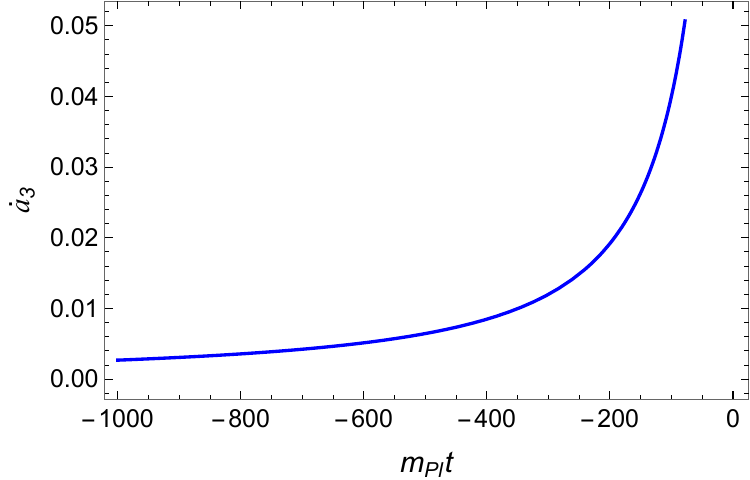}}
\caption{(a) Functions $a_i(t)$, $a(t)$ and $\dot{a}_3$ with the initial conditions    $p_1 = 10^3$, $p_2 = 2\times 10^3$,  $p_3 = 3\times 10^3$,  $c_1 = -0.3$, $c_2 = -0.2$,  $\rho_0 = 3.55\times 10^{-5}\; m_{\text{Pl}}$    for the Bianchi I universe filled with a radiation fluid  $\rho_M=\rho_0/v^{4/3}$ \cite{Motaharfar:2026jcj}.
    Initial conditions are imposed at $t = 0$. (b) The corresponding plot of $\dot{a}_3(t)$ before bounce, $t < 0$. }
    \label{MS_Fig5}
\end{figure} 
%%%%%%%%%%%%%%%%%%%%%%%%%%%%%%%%%%%%%%%%%%

Recently,  the authors in ~\cite{Motaharfar:2026jcj} questioned the above conclusion by considering numerically some particular initial conditions. In addition, they also argued that the post-bounce universe never becomes truly classical. In this  {Note}, we first examine  their analysis and underlying assumptions, and find that the special cases considered by them are not generic; rather, generic initial conditions do lead to quantum damping of the cosmological shear. This damping happens well within the quantum regime $t \approx {\cal{O}}\left(t_{\text{Pl}}\right)$, during which the classical limit conditions 
$$
\left|\mu_ic_i\right| \ll 1,
$$
are not held yet, as stated explicitly in \cite{Gan:2025uvt}. In fact, it is precisely this that shows clearly that the damping mechanism is quantum in nature.  Furthermore, we discuss the mechanism by which a classical regime can emerge in the post-bounce region through the super-Hubble backreaction first discovered in ~\cite{Tsamis:1992sx,Tsamis:1994ca,Mukhanov:1996ak,Tsamis:1996qm} and then  analyzed in great details in~\cite{Marozzi:2006ky,Abramo:2001dd,Losic:2005vg,Brandenberger:2018fdd,Woodard:2025smz,Glavan:2026pug}. The backreaction of super-Hubble fluctuations effectively contributes an energy–momentum tensor, which behaves as a negative cosmological constant, i.e., a component with negative energy density and equation of state $\omega=-1$. As the phase space of super-Hubble modes grows exponentially  {[cf. Fig. \ref{New_HH}],} this effective negative contribution builds up, grows and decreases the initially effective $\Lambda$, whereby a classical universe is achieved. In this way, the effective cosmological constant is dynamically reduced through a self-adjusting mechanism \cite{Brandenberger:1999su}. This contribution increases during the accelerated expansion and decreases as the quasi-de Sitter phase ends, tied  naturally  to the background dynamics. Non-perturbative calculations have also been performed indicating that the effect can be sufficiently large to significantly reduce an initially large $\Lambda$, see for instance \cite{Brandenberger:2018fdd}. Further studies of this mechanism will be reported on another occasion   \cite{Gan:2026}.  

The rest of this Note is organized as follows. In Section \ref{SecII},  we  address the issue of the genericness of the quantum damping of the cosmological shear discovered in \cite{Gan:2025uvt}, while in Section \ref{SecIII} we outline  a plausible post-bounce mechanism for the onset of classicalization. In Section \ref{SecIV}, we summarize our main conclusions and provide some discussing remarks.

%%%%%%%%%%%%%%%%%%%%%%%%%%%%%%%%%%%%%%%%%
\section{Genericness of Quantum Damping of Cosmological Shear}
\lb{SecII}

To show our above claims, we first note, within the standard relational framework used in LQC/mLQCs, that the vacuum case considered in \cite{Motaharfar:2026jcj}  is not well-defined, where matter fields serve as relational clocks that define physical evolution. In the absence of such a clock, the Hamiltonian constraint reduces to a timeless difference equation, and no meaningful notion of dynamics can be extracted. Therefore, conclusions drawn from vacuum solutions are not directly comparable to the matter-coupled scenarios relevant for cosmology and must be interpreted with caution. To illustrate this, one can simply consider  a massless scalar field $\phi$ in LQC, for which the Schr\"odinger-like equation takes the form \cite{Ashtekar:2006wn}
\bqn
\lb{eq1}
\partial^2_{\phi}\Psi(v, \phi) = - \Theta  \Psi(v, \phi),
\eqn
where $v$ denotes the three-volume  
of the flat Friedmann-Lema\^itre-Robertson-Walker (FLRW) universe, $ds^2 = - dt^2 + a(t)d^2\vec{x}$,  and $\Theta$ a Hermitian difference operator  that connects the wavefunction $\Psi(v, \phi)$ with the ones $\Psi(v\pm 4, \phi)$ given at the 3-volumes
$v\pm 4$ for a given moment $\phi$, reflecting the fact that the three-dimensional spatial space is now  discretized. Equation~\eqref{eq1} clearly shows  the importance of the scalar field. Without it, the Hamiltonian constraint 
$$
\left(\hat{\cal{H}}_g + \hat{\cal{H}}_m\right) \Psi(v, \phi) = 0,
$$
simply leads to the difference equation $\Theta  \Psi(v, \phi) = 0$. Then, the physical meaning of this difference equation without time-evolution is unclear. The same argument is equally applicable to other spacetimes, including the Bianchi I universe
\bqn
\lb{eq2}
ds^2   = - dt^2 + \sum_{i = 1}^{3}{a_i^{2}(t) \left(dx^i\right)^2},
\eqn
where   
$a_i(t)$ are the scale factors along the three spatial directions. 
Therefore, in the rest of this paper, we discard the vacuum case,  despite the fact that most of our arguments given below can also be applied to this case.

Adopting the Ashtekar variables~\cite{Ashtekar:1986yd}, the momenta are defined as \footnote{In this paper, we focus on the effective dynamics of the Bianchi I universe for 
sharply peaked trajectories, such that $a_i \ge 0$ \cite{Ashtekar:2009vc}.}
$p_i \equiv  a_ja_k \; (i \not= j\not=k)$. 
Then, the effective gravitational Hamiltonian ${\cal{H}}_{\text{eff}}^{\text{mLQC-I}}$ 
for the Bianchi I universe in the framework of mLQC-I was derived  
in~\cite{Garcia-Quismondo:2019kav,Garcia-Quismondo:2019dwa}  and the corresponding Hamiltonian equations for ${p}_i$ and ${c}_i$ are given explicitly in
\cite{Gan:2025uvt}, where $c_i$ denotes the connection component conjugate to $p_i$,
satisfying the canonical relations $\left\{c_i, p_j\right\} = 8\pi G \gamma \delta_{ij}$, where $\gamma$ is the Barbero-Immirzi parameter
(in all of our numerical results, we choose $\gamma \approx 0.2375$, 
as suggested from black hole thermodynamics in LQG~\cite{Meissner:2004ju}). 

In~\cite{Motaharfar:2026jcj} the authors considered numerically a particular case with the initial conditions:  $p_1 = 10^3$, $p_2 = 2\times 10^3$,  $p_3 = 3\times 10^3$, $c_1 = -0.3$, $c_2 = -0.2$.  {Note that the initial time can be always set to $t = t_i = 0$ by using the translation symmetry, $t \rightarrow t + t_0$, of the Bianchi universe, without loss of generality. Therefore, in the rest of this paper  we shall adopt this assumption without any further explanations.}  The corresponding evolutions  of $a_i$ and $a \left(\equiv (a_1 a_2 a_3)^{1/3}\right)$ were shown in Fig. 5 of Ref. \cite{Motaharfar:2026jcj}, which is re-produced currently in Fig.~\ref{MS_Fig5} with a larger range of $t/t_{\text{Pl}} \in \left(-10^3, 10^3\right)$, so the asymptotic behavior of the solution can be seen clearly. In the range 
$t/t_{\text{Pl}} \in \left(0, 80\right)$, our results are the same as those presented in \cite{Motaharfar:2026jcj} (within the numerical errors). Then, our comments are in order: 

\begin{itemize}

\item First, we clarify what is meant by generic initial conditions in the present context. 
 {In this work, we define physically admissible initial data as those corresponding to a genuinely collapsing Bianchi I universe, in which all three directional Hubble rates are initially negative, i.e., $c_i(0) < 0$ for $i=1,2,3$. Since the directional Hubble rates $H_i$ are proportional to combinations of $c_i$'s in the effective dynamics, the condition $c_i(0) <0$ corresponds to contraction in all spatial directions initially.
This ensures that the spacetime undergoes a true three-dimensional contraction prior to the bounce and can evolve into a macroscopic $(3+1)$-dimensional universe after the bounce. Configurations in which one or more directions are initially expanding, or remain permanently at the Planck scale, do not satisfy this requirement and are not representative of viable cosmological histories.}

 {The specific examples analyzed in \cite{Motaharfar:2026jcj} are not generic. In particular, solving the Hamiltonian constraint with their initial data yields $c_3(0) > 0$, implying that the spacetime is already expanding along one direction in the pre-bounce phase, as shown explcitly in Fig. \ref{MS_Fig5} (b), which was also noticed in \cite{Motaharfar:2026jcj}. As a consequence, the resulting evolution corresponds to a mixed expanding–contracting configuration rather than a genuine collapse. Furthermore, as shown in Fig.~\ref{MS_Fig5} (a), one spatial direction remains at the Planck scale throughout the evolution, effectively leading to a lower-dimensional universe. We therefore conclude that such configurations lie outside the physically relevant sector of the phase space.}

\item  Second, in contrast to \cite{Motaharfar:2026jcj}, when the initial conditions satisfy $c_i(0) < 0$ for all $i$'s, we find that the post-bounce evolution exhibits a universal behavior characterized by exponential expansion in all three spatial directions. This is illustrated in Fig.~\ref{New_Fig5}, where we adopt the same initial values for $p_i$, $c_1$, and $c_2$ as in \cite{Motaharfar:2026jcj}, but now choose $c_3 = -6.23\times 10^{-9} < 0$. The resulting solution shows that all scale factors rapidly approach the form
\bqn
\lb{eq3a}
a_{i}(t) = a_i^{(B)} e^{A (t-t_B)},  \; (i = 1, 2, 3),
\eqn
for $ t > t_B$,  where 
\bq
\lb{eq3}
A \equiv \frac{3}{\left(1+\gamma^2\right) \sqrt{\Delta}} \simeq \frac{1}{0.8 t_{\text{Pl}}},
\eq
$\Delta \left(\equiv 4\pi \sqrt{3} \gamma \ell_{\text{Pl}}^2\right)$ denotes the non-zero minimal area gap of LQG~\cite{Ashtekar:2004eh,Thiemann:2007pyv,Rovelli:2014ssa}, and $t_B$ denotes the bounce time, while $a_i^{(B)}$ are the corresponding amplitudes of $a_i(t)$ at the bounce. 
From Fig.~\ref{New_Fig5}, it can be seen clearly that right after the bounce the expansion factors soon expand exponentially as given by Eqs.~\eqref{eq3a} and~\eqref{eq3}. With different choices of initial conditions and matter fields, we find that the behavior illustrated in Fig.~\ref{New_Fig5}  is universal and is not only independent of the choice of the initial conditions (as long as $c_i(0) < 0$, so initially the spacetime is collapsing in all three spatial directions), but also independent of matter fields, dust, radiation, scalar fields with potential, and so on.  

%%%%%%%%%%%%%%%%%%%%%%%%%%%%%%%%%%%%%%%%%%
\begin{figure}[h!]  
\includegraphics[width=0.47\textwidth]{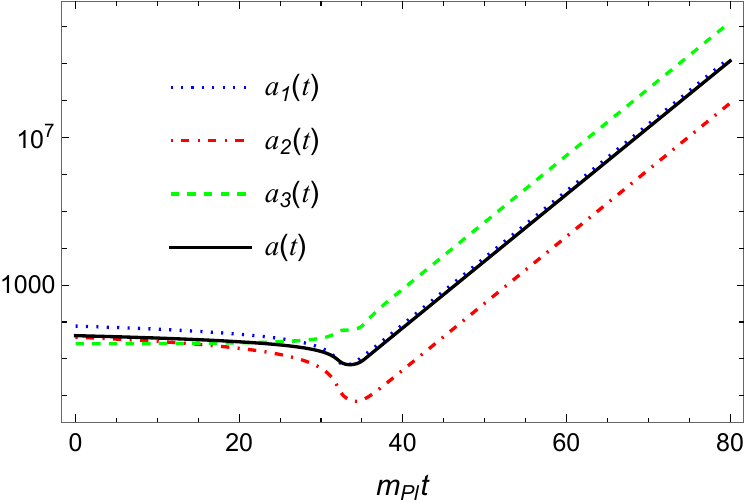}
\caption{Functions $a_i(t)$ and $a(t)$ with the same initial conditions for $p_i$, $c_1$ and $c_2$, as those chosen in Fig.~\ref{MS_Fig5}, but now with   $c_3 = - 6.23\times 10^{-9}$     for the Bianchi I universe filled with a radiation fluid.
    Initial conditions are imposed at $t = 0$.  }
    \label{New_Fig5}
\end{figure} 
%%%%%%%%%%%%%%%%%%%%%%%%%%%%%%%%%%%%%%%%%%

\item  {Third,} the above numerical results  are strongly supported by perturbation analyses, in which the Bianchi I universe is considered as linear perturbations of  the flat FLRW universe
\bqn
\lb{eq4}
a_i(t) = a(t) e^{\theta_i(t)}, \quad \sum_{i = 1}^{3}{\theta_i} = 0, \quad \left|\theta_i\right| \ll 1.
\eqn 
Then,  {focusing ourselves in the post-bounce region where $\dot{v} > 0$, we find that} the effective Hamiltonian equations for ${p}_i$ and ${c}_i$ lead to  
\cite{Gan:2025uvt}
\bqn
\label{eq5}
     \ddot{\theta_i}+\mathcal{F}(t)\dot{\theta_i}=0,\quad
    \mathcal{F}(t) \simeq A+ \frac{B}{a(t)^{3 (\omega +1)}},
\eqn
where $A$ is given by Eq.~\eqref{eq3}, $\omega$ denotes the equation of state of the matter field,
$P = \omega \rho = \omega \rho_0/v^{1+\omega}$, and
\bqn  
\lb{eq7}
B \equiv \frac{80 \pi  \gamma ^2 G \sqrt{\Delta}  \rho_0 \omega  }{4 \gamma ^2+1}.
\eqn
Therefore, as long as $\omega > -1$, we have $\mathcal{F}(t) \approx A$. Hence, Eq.~\eqref{eq5} has the general  solutions
\bqn
\lb{eq8}
\theta_i(t) \simeq \theta^{(0)}_i e^{- A t} + \theta^{(1)}_i,
\eqn
where $\theta^{(n)}_i$'s are integration constants, and $\theta^{(1)}_i$ can be sent to zero by rescaling, $x^i \rightarrow e^{\theta^{(1)}_i} x^i$.  From Eq.~\eqref{eq8} we can see that the anisotropy $\theta_i$ decreases exponentially
and becomes very small as $t \gg A^{-1} \simeq 0.8 t_{\text{Pl}}$, which is well within the quantum regime.  
More importantly, this result is not restricted to a particular choice of initial conditions nor a specific matter field, as long as it satisfies the weak energy condition, $\omega > -1$.

\end{itemize}

Therefore, while special configurations outside the class defined above may exhibit non-isotropizing behavior, such cases do not correspond to physically relevant collapsing cosmological scenarios. The conclusions of \cite{Motaharfar:2026jcj} are thus not in contradiction with our results, but rather pertain to a different sector of the phase space that is not relevant for realistic cosmological scenarios.

%%%%%%%%%%%%%%%%%%%%%%%%%%%%%%%%%%%%%%%%%%
\begin{figure}[h!]  
\includegraphics[width=0.47\textwidth]{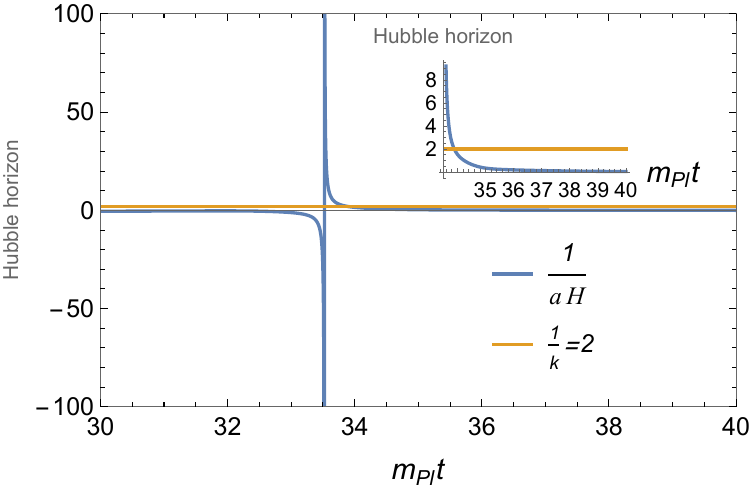}
\caption{The Hubble horizon $L_H \equiv (aH)^{-1}$ with the same initial conditions for $p_i(0)$ and $c_i(0)$, as those chosen in Fig.~\ref{New_Fig5}   for the Bianchi I universe filled with a radiation fluid. 
    Initial conditions are imposed at $t = 0$, and  qualitative behavior of the Hubble horizon is quite similar for other matter fields.   {All quantities in the figure are given in Planck units.}}
    \label{New_HH}
\end{figure} 

%%%%%%%%%%%%%%%%%%%%%%%%%%%%%%%%%%%%%%%
\section{ Perturbations of Super-Hubble Modes and Classicalization}
\lb{SecIII}

Now let us turn to the second question raised in \cite{Motaharfar:2026jcj}, regarding   smoothly transforming such a de Sitter phase with a Planck-size cosmological constant to a classical one in the post-bounce region. 

Let us first note that
this accelerated stage emerges 
dynamically from quantum geometry rather than from a fundamental cosmological constant, 
it need not persist indefinitely. 
 {As the exponential damping of anisotropies demonstrated above occurs entirely within the quantum regime, where $|\mu_i c_i| \gtrsim 1$, and therefore does not rely on the existence of a classical limit. The question of how a classical universe subsequently emerges is conceptually distinct from the mechanism responsible for shear suppression.}
A plausible mechanism for the dynamical emergence of a classical regime is provided by the backreaction of super-Hubble modes, which has been extensively studied in the context of de Sitter spacetime~\cite{Tsamis:1992sx,Tsamis:1994ca,Mukhanov:1996ak,Marozzi:2006ky,Abramo:2001dd,Losic:2005vg,Brandenberger:2018fdd,Woodard:2025smz,Glavan:2026pug}. 
In such scenarios, the cumulative effect of the super-Hubble modes  can effectively reduce the expansion rate, leading to a gradual relaxation of an initially large effective cosmological constant, allowing radiation 
or matter to dominate and classical evolution to resume. Such cumulative effects
become more efficient at the Planck energy \cite{Marozzi:2006ky}. This can be understood as follows: Quantum fluctuations are stretched beyond the Hubble horizon rapidly after the bounce, where the Hubble horizon behaves as 
\bq
\lb{eq8b}
L_H \equiv \frac{1}{aH} = \begin{cases}
    \infty & t = t_B, \cr
    \frac{e^{-H(t-t_B)}}{a_B H}, & t \gtrsim {\cal{O}}\left(t_{\text{Pl}}\right), \cr
\end{cases}
\eq
where $a_B$ is the value of the average expansion factor at the bounce and $H = A = {\cal{O}}\left(m_{\text{Pl}}\right)$, with $A$ given by Eq.~(\ref{eq3}). Thus, for a given comoving wavenumber (momentum) $k$, it will soon cross the Hubble horizon at $t_k$, as shown by the horizontal solid orange line in Fig. \ref{New_HH},  become a super-Hubble mode, freeze out, and squeeze (here $t_k$ is given by $k = aH(t_k)$). The fact that new modes are kept injecting into the super-Hubble phase space is crucial for the backreaction to be able to cancel the bare cosmological constant effectively and finally lead the accelerated phase to an end. In the current case, such a phase space is built up exponentially fast  with  a   rate $H \approx {\cal{O}}\left(m_{\text{Pl}}\right)$, which might be the highest rate that we could physically have.

To be more specific, let us first note that  many works in the literature have supported the claim that  in de Sitter space,   
the backreaction of the super-Hubble fluctuations can effectively 
reduce the expansion rate~\cite{Tsamis:1992sx,Tsamis:1994ca,Mukhanov:1996ak}.  
The effective Hubble parameter can be cast in the 
form~\cite{Abramo:2001dd}
\begin{equation}
\lb{eq9}
H_{\rm eff} = H \left[ 1 - {\cal D}(Ht)^n + \cdots \right],
\end{equation}
where 
${\cal{D}}$ and $n$ are two parameters that characterize radiation. 
For example, 
when the two-loop effects of gravitons were taken into account,  ${\cal{D}} \simeq \left(G\Lambda/3\pi\right)^2/6$ and $n = 2$ \cite{Tsamis:1996qm}, while for a scalar field with quartic self-interaction  $V(\phi)=\lambda \phi^4/4!$, one has ${\cal{D}} = \lambda^2 G\Lambda/(2^73^4\pi^5)$ and $n=4$. 

In addition, recently it was also shown that even when non-linear perturbations are taken into account, similar conclusions are obtained~\cite{Brandenberger:2018fdd,Woodard:2025smz,Glavan:2026pug}, which further suggest that the back reaction of the super-Hubble perturbations is real and relevant.  In particular, in \cite{Brandenberger:2018fdd} it was shown how the expansion rate measured by a (secondary) clock field obtains a negative contribution from fluctuations, a contribution whose absolute value increases in time, i.e. $\Delta H_{\text{eff}}<0$ and $d(\Delta H_{\text{eff}})/dt<0$.  This analysis did not make use of any perturbative
expansion in the amplitude of the inhomogeneities, supporting the conclusion that the super-Hubble fluctuations lead to a dynamical relaxation of the cosmological constant. Taken together, all these results suggest that for $\Lambda \approx {\cal{O}}\left(m_{\text{Pl}}^2\right)$,  the expansion rate $H_{\rm eff}$ can  be very efficiently reduced. 
 As this mechanism only depends on modes that become super-Hubble and backreacts on the background dynamics, it comes with no cost as it is intrinsic and effective for Planck size cosmological constants.

%%%%%%%%%%%%%%%%%%%%%%%%%%%%%%%%%
\section{Concluding Remarks}
\lb{SecIV}

In this brief Note, we  have revisited the claims raised recently in~\cite{Motaharfar:2026jcj},
by carefully examining the underlying assumptions and the class of initial conditions considered there. We have shown explicitly that the examples analyzed 
in~\cite{Motaharfar:2026jcj} correspond to configurations that do not represent physically admissible contracting universe, as they involve mixed expanding–contracting directions and lead to effectively lower-dimensional post-bounce geometries. Restricting to physically relevant initial conditions corresponding to genuine three-dimensional contraction, we have shown that the quantum damping of cosmological shear is a robust dynamical feature.
This conclusion is supported first by numerical analysis carried out in Section \ref{SecII} and then by perturbative analysis in Section \ref{SecIII}, in which we have demonstrated that the post-bounce evolution admits an isotropic attractor, with anisotropies decaying exponentially and independently of the matter content, provided that the weak energy condition is satisfied.

Regarding  to the question how one can smoothly transform such a de Sitter phase with a Planck-size cosmological constant to a classical one in the post-bounce region,
 in Section \ref{SecIV} we have shown that existing results in the literature suggest that it can operate efficiently on timescales of order $t_{\rm Pl}$ when the effective cosmological constant is Planckian, while a quantitative implementation of the mechanism described above within the present mLQC-I framework is being developed~\cite{Gan:2026}. This indicates that, despite the fact that  the transition was not made explicit  in the specific examples considered in \cite{Motaharfar:2026jcj}, a classical regime can nevertheless emerge dynamically in physically relevant models presented  in \cite{Gan:2025uvt}.

%%%%%%%%%%%%%%%%%%%%%%%%%%%%%%%%%%%%%%%%
\acknowledgements
A.W. would like to   thank Prof. Paramapreet Singh for valuable discussions and comments. W.-C. G. is supported by the National
Natural Science Foundation of China under the Grant No. 12405064, Jiangxi Provincial 
Natural Science Foundation under the Grant No. 20242BCE50055, and the Initial 
Research Foundation of Jiangxi Normal University.  L.L.G is supported by research 
grants from Conselho Nacional de Desenvolvimento Cientıfico e Tecnologico (CNPq),
Grant No. 307636/2023-2 and from the Fundacao Carlos Chagas Filho de Amparo a Pesquisa do Estado do Rio de
Janeiro (FAPERJ), Grant No. E-26/204.598/2024. 
R.O.R. is partially supported by research grants from CNPq, Grant No. 307286/2021-5, and FAPERJ, Grant
No. E-26/200.415/2026.
A.W. is partially supported by the US NSF  grant, PHY-2308845.
  
%%%%%%%%%%%%%%%%%%%%%%%%%%%%%%%%%%%%%%%%%%%%%  

%%%%%%%%%%%%%%%%%%%%%%%%%%%%%%%%%%%%%%%%%%%%%%

\end{document}